\documentclass[10pt, twocolumn]{IEEEtran}

\usepackage{MathDefs}
\usepackage{balance}

\begin{document}

\title{A Novel Approach for Resistance Estimation from the HPPC Test}
\title{An Improved Approach to Estimate the Internal Resistance of a Battery During the HPPC Test}

\author{Prarthana Pillai$^{\star \dagger}$,
Smeet Desai$^{\dagger}$,
Krishna R. Pattipati$^{\ddagger},$  and
Balakumar Balasingam$^{\dagger}$
\thanks{Submitted to {\em IEEE Journal of Emerging and Selected Topics in Industrial Electronics}, May 2025.}
\thanks{$^\star$Prarthana Pillai is the corresponding author at pillaip@uwindsor.ca.}
\thanks{$^\dagger$The authors are with the Department of Electrical and Computer Engineering, University of Windsor, 401 Sunset Ave., Windsor, ON N9B3P4, Canada, TP: +1(519)-253-3000 ext. 5431, E-mail: \{pillaip, desai56, singam\}@uwindsor.ca.}
\thanks{$^\ddagger$The author is with the Department of Electrical and Computer Engineering, University of Connecticut, 371 Fairfield Way, Storrs, CT 06269, USA, TP: +1(860)-486-2890, E-mail: krishna.pattipati@uconn.edu.}
}

\maketitle

\begin{abstract}
This paper considers the problem of resistance estimation in electronic systems including battery management systems (BMS) and battery chargers. In typical applications, the battery resistance is obtained through an approximate method computed as the ratio of the voltage difference to the applied current excitation pulse or vice versa for admittance. When estimating the battery resistance, this approach ignores the change in the open circuit voltage (OCV) as a result of the excitation signal. In this paper, we formally demonstrate and quantify the effect of the OCV drop on the errors in internal resistance estimation. Then, we propose a novel method to accurately estimate the internal resistance by accounting for the change in OCV caused by the applied current excitation signal. The proposed approach is based on a novel observation model that allows one to estimate the effect of OCV without requiring any additional information, such as the state of charge (SOC), parameters of the OCV-SOC curve, and the battery capacity. As such, the proposed approach is independent of the battery chemistry, size, age, and the ambient temperature. A performance analysis of the proposed approach using the battery simulator shows significant performance gain in the range of 30\% to more than 250\% in percentage estimation error. Then, the proposed approach is applied for resistance estimation during the hybrid pulse power characterization (HPPC) of cylindrical Li-ion battery cells. Results from tested batteries show that the proposed approach reduced the overestimated internal resistance of the batteries by up to 20 m$\Omega$.
\end{abstract}  

\begin{IEEEkeywords}
HPPC Test,
Resistance estimation,
Li-ion batteries, 
state of charge
\end{IEEEkeywords}


\section{Introduction}

Electric vehicles (EVs) employ lithium-ion batteries due to their high energy density \cite{zhang2022status} and low rate of self-discharge \cite{capasso2014experimental}. 
Battery management system (BMS) plays a critical role in the reliable operation of batteries in EV \cite{balasingam2023robust}. 
The BMS controls the voltage limits of the battery, provides protection from overcharging and over-discharging, and estimates crucial parameters like state of charge (SOC), state of health (SOH), and state of power (SOP) of the battery \cite{rahimi2013battery}. 
Open circuit voltage (OCV) and internal resistance of the battery are crucial for estimating the SOC, SOH, and SOP, and must be accurately estimated for effective battery management.   

Electrical resistance is defined as the property of any material describing its opposition to the flow of electrical current. 
The resistance of any material is defined using Ohm’s law, which states that resistance is the ratio of the voltage to the current. 
The current and voltage quantities can be obtained by controlled injection of one and the measurement of the other, i.e., voltage can be applied and current can be measured (potentiostatic measurements) or current can be applied and voltage can be measured as response (galvanostatic measurements). 
It is a fundamental quantity used in circuit design across multiple industries for its usage as a measure of material property, quality, reliability and safety. 

In material sciences, resistance measurement is used for insulation and damage detection \cite{ saberi2021new}. 
In the medical field, impedance measurement is useful for biomedical devices and live tissues. 
Bioimpedance measurement involving live tissues follows stringent standards to maintain current ranges between 100mA  and 100$\mu$A \cite{kassanos2018electrical}, 
with further restrictions for heart patients \cite{rubio2020analysis}. 
Accurate impedance monitoring is also crucial for motor devices, ear implants, and cell monitoring to prevent harmful voltage levels \cite{kassanos2021bioimpedance}.

In the automotive industry, battery resistance estimation has important applications.
One such example is in the estimation of the SOH of a starter battery of the internal combustion engines. 
These batteries are typically made of lead-acid chemistry and measure either 6V or 12V. 
To measure their SOH, a high current load (typically 100A) is applied to the battery and the voltage drop across the terminals is measured \cite{SchumacherBT100, PerformanceToolW2988, Bosch100Amps, Neiko40510A, CenTech100Amp}.
The amount of voltage drop serves as a measure of the SOH; here, the SOH is a combined measure of both the power fade and capacity fade of the starter battery. 
In battery electric vehicles, the battery pack powers the entire vehicle operation.
As such, more detailed battery health monitoring is necessary; here, the power fade and capacity fade are separately monitored.  
Other circuit diagnostics in electrical energy storage systems are performed using a variety of instruments, varying in their precision and measurement capabilities. Some instruments include multimeters, Ohmmeters, Wheatstone and Kelvin Bridges, micro and meg-ohmmeters, RLC meters \cite{keysight_resistance_guide}. In many of these measurement systems for batteries, a single value or frequency-dependent resistance measurement is obtained in the steady-state, directly or indirectly using the measured voltage drop. 

The focus of the present work is towards improving the accuracy of resistance measurement in batteries and other similar energy storage systems. 
Electrical energy storage systems experience a voltage bias when controlled voltage/current is injected for the resistance/impedance measurement. 
This means that the current/voltage recorded in response to the supplied injection is affected by this bias and this bias affects the computed resistance.

The voltage bias recorded in the response of the battery to injected current/voltage is attributed to its complex electrochemical and nonlinear behaviour. 
The non-linear battery behavior is modelled using electrical equivalent circuit models (ECM) consisting of resistive, capacitive, and inductive components. 
The values of the ECM components determine the dynamic behaviour of the battery. Further, ECM modelling is crucial for the management of BMS functionalities.
An evaluation of 14 ECMs is done in \cite{jung2023characteristics}, detailing their advantages and complexities. 
High-fidelity ECMs may accurately represent the battery behavior; however, estimating their parameters may be challenging. 
On the other hand, the parameters of low-fidelity ECMs can be easily estimated \cite{ospina2021comparison}. 
Based on the domain of estimation, existing ECM parameter identification can be categorized into frequency, time-domain and standardized characterization methods.

Frequency-domain approaches employ sinusoidal signals of frequencies ranging from few mHz to several kHz \cite{meddings2020application} to obtain the parameters of the selected ECM of a battery. 
Electrochemical Impedance Spectroscopy (EIS) is a widely used method in BMS research \cite{westerhoff2016electrochemical}. 
In EIS, a small amplitude current or voltage of varying frequencies is applied to the battery \cite{martinez2020eis};
the measured response of the battery is converted into the frequency domain using discrete Fourier transform \cite{orazem2008electrochemical} and the impedance spectrum is obtained in the frequency domain, later used by algorithms for the estimation of ECM parameters \cite{wu2023battery,pizarro2021ga,nunes2024impedance}.
Even though EIS is the most widely used method for impedance estimation, it is disadvantaged by the fact that it requires a long measurement time. Also, EIS approaches require high-precision hardware capable of data measurement at a wide range of frequencies \cite{novakova2024review}.

Time-domain approaches use a pulse current to measure the instantaneous voltage response of the battery. 
Similar to the frequency-domain approach, the parameters of selected ECM are captured using various algorithms in an online or offline manner. 
Online parameter estimation updates the parameters throughout the operation, while offline parameter estimation is done on data accumulated after the operation. 
Extraction of the ECM parameters is done through variants of Kalman filter \cite{he2011state, he2013state, zhu2019state} and linear parameter extraction methods \cite{li2015state, zhang2018online, sun2021state, tan2021online}.
Based on the pre-selected model to define the internal physico-chemical processes of the battery in time-domain approaches, significant changes to the overall computational demand and accuracy are expected. 
Further, the non-linear nature of the battery requires the use of complex optimization algorithms to solve the ECM parameter identification problem \cite{zhao2022comparative}.

Two major issues with both the frequency-domain and time-domain approaches reviewed so far are that they require precise measurement systems and extensive computing power to compute the high-fidelity ECM parameters. Many practical applications utilize standardized approaches that are easy to implement and require less hardware and computational resources. 

The Hybrid Pulse Power Characterization (HPPC) test is one such method and is used to quantify the dynamic power capability of a battery, which is its ability to store and deliver power during charging and discharging. The HPPC test is also a standardized approach to data acquisition to estimate the ECM parameters \cite{li2020investigation}. HPPC test is used to accumulate data for ECM parameter estimation in \cite{he2011state}, \cite{li2015state},  \cite{mahboubi2022state}, \cite{bi2020adaptive}, \cite{zhang2016improved}, \cite{pang2019parameter}, \cite{bialon2023hppc}, and \cite{zhang2022effect}. A variation of HPPC pulse is observed in the literature, where the length of discharge pulse and shape of charge pulse varies \cite{zhao2024modified}, \cite{jiang2021state}, \cite{barai2018study}, \cite{lee2015maximum},  \cite{shh1999pngv}. 
The charge and discharge resistance are calculated as the slope of voltage and current. The calculated resistances are further used to calculate the pulse power capability, also referred to as SOP \cite{pillai2024sop} of the battery, at different SOCs. The results of the HPPC test can be scaled from cell level to pack level, which makes it a preferable test for data collection.

In this paper, the charge-discharge HPPC pulse \cite{barai2018study, christophersen2015battery} is considered for the demonstration of the proposed approach. The duration of the current pulse in HPPC is varied to assess the effect of various electrochemical processes being activated within the battery. Typical current pulse durations are 0.01, 2, 5 and 10 seconds, where the resistance computed using the shorter duration represents the Ohmic resistance, whereas the resistance computed using longer pulses represents the resistive effect of the battery due to the polarization effect caused by the electrochemical processes within the battery. This standardized approach offers simplicity to gauge the response of the battery against nonlinear modeling approaches that require elaborate measurement and computing systems to estimate the parameters of the polarization effect \cite{sihvo2020novel, gong2022ev, islam2018circuit, ahmed2014reduced, pillai2022optimizing}.  

It is noted, however, that the existing approach computes the resistance as a ratio of the voltage difference and the applied current. Here, it is assumed that the change in the battery OCV, due to the applied current, is negligible. This assumption is appropriate when the pulse duration is short and the magnitude of the current pulse is relatively small. As the pulse duration and the magnitude of the current pulse increase, the change in OCV due to the applied pulse becomes significant enough to cause noticeable change in the estimated resistance. It is shown in this paper that ignoring the OCV drop in resistance estimation using the HPPC standard leads to {\em overestimation of the resistance}. Therefore, the present work aims to accurately predict the impedance using a novel method of resistance estimation from the HPPC test data. The proposed method is designed to estimate the OCV drop and account for it in the resistance estimation of the battery. The accuracy of the modified estimation approach is demonstrated on the HPPC test data collected from a battery simulator using MATLAB and data collected from four real cells.

The remaining sections of this paper are organized as follows:
Section \ref{sec:hppctest} describes the HPPC test procedure and the process of calculating the resistance from the HPPC test.  
Section \ref{sec:problemDef}  presents the limitations of existing approaches. 
Section \ref{sec:proposed} delineates the proposed novel approach to resistance and OCV estimation. 
Section \ref{sec:simresults} contrasts the existing and proposed approaches for resistance estimation.
Section \ref{sec:data-collection} presents the analysis of real data using the existing and proposed approaches, and the paper is concluded in Section \ref{sec:conclusion}.

 \section{HPPC Test} 
\label{sec:hppctest}

The HPPC characterization is performed within the manufacturer's specified voltage limits, i.e., between $V_{\rm max}$, the maximum allowable voltage, and $V_{\rm min}$, the minimum allowable voltage. 
The magnitude of the discharge current for the HPPC discharge profile determines the type of test, and consequently, the power expended. In the {\em low-current HPPC test}, a current of at least $C$ amperes is applied, where $C$ denotes the battery capacity in Ah. The {\em high-current HPPC test} uses a current of $0.75 I_{\rm max}$ amperes for high-discharge power computation, where $I_{\rm max}$ is a manufacturer-specified absolute maximum allowable pulse current.

\subsection{HPPC pulse}
\label{sec:hppcpulsedesc}
\begin{figure}[h!]
\begin{center}
{\includegraphics[width= 1\columnwidth]{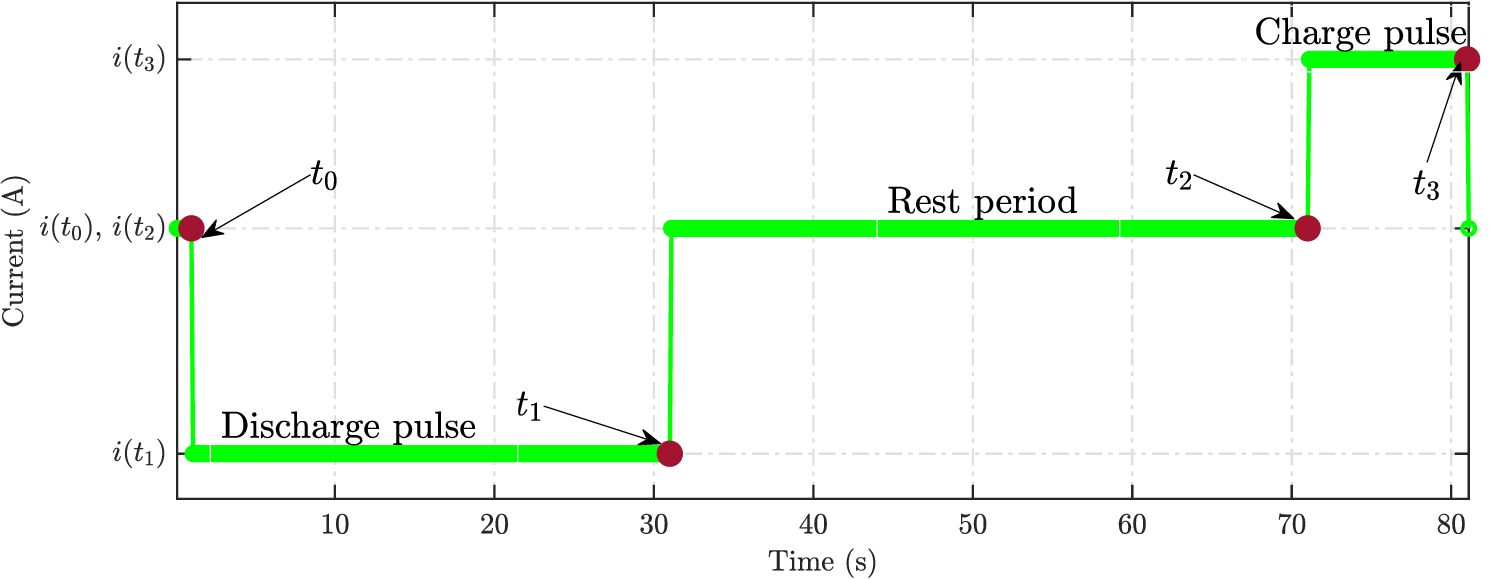}} 
\caption{Standard HPPC pulse \cite{christophersen2015battery}.\label{fig:HPPC_Pulse_sim}}
\end{center}
\end{figure}

The HPPC pulse profile standardized in \cite{christophersen2015battery} is 80 seconds long as shown in Fig. \ref{fig:HPPC_Pulse_sim}. 

Each pulse consists of three regions: (i) Discharge Pulse. This pulse is the first region between time instances $t_0$ and $t_1$ where a discharge current $I_{\rm dis}$ is applied for 30 seconds. The magnitude of  $I_{\rm dis}$ is decided based on the low-current or high-current HPPC test to be performed.
(ii) Rest period. The battery is rested for 40 seconds between times $t_1$ and $t_2$.
(iii) Charge (Regenerative) Pulse. Here, a charging current  $I_{\rm ch}$ of magnitude 0.75$\times I_{\rm dis}$ is applied for 10 seconds between times $t_2$ and $t_3$. 

The corresponding voltages for discharge pulse between times $t_0$ and $t_1$ are $v (t_0)$ and $v(t_1)$, and the corresponding currents are $i(t_0)$ and $i(t_1)$, respectively. Similarly, for the charge pulse times between $t_2$ and $t_3$, the corresponding voltages are $v (t_2)$ and $v (t_3)$, and currents are $i(t_2)$ and $i(t_3)$, respectively. It should be noted that the current values in Fig.\ref{fig:HPPC_Pulse_sim} are relative and may vary depending on the type of HPPC test to be performed as mentioned earlier in Section \ref{sec:hppctest}.

\subsection{HPPC test methodology}
The HPPC test starts with a fully charged battery, at terminal voltage, $V_{\rm max}$. Prior to HPPC test, a standard initiation procedure is applied to the battery. 

\begin{figure}[h!]
\begin{center}
{\includegraphics[width= 1\columnwidth]{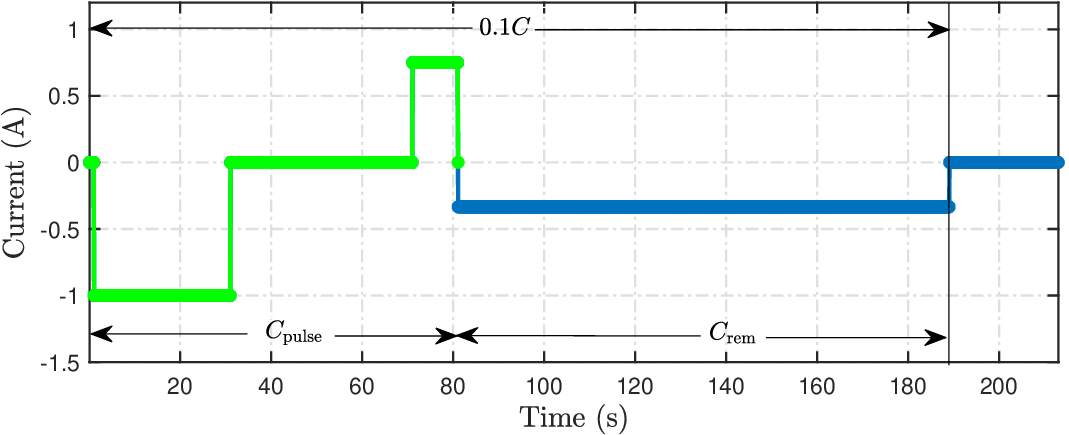}} 
\caption{Current profile for HPPC pulse and discharge to next SOC. Total coulombs to be removed is $(0.1)C \, {\rm Ah}$ which is a combination of $C_{\rm pulse}$ and $C_{\rm rem}$.\label{fig:capdef}}	
\end{center}
\end{figure}

First, the battery is discharged with a current of $C/3$ amperes until the voltage reaches $V_{\rm min}$. The battery is rested for a standard rest period (typically 1 hour). Following the rest, the battery is charged to $V_{\rm max}$ using the standard charging current specified by the manufacturer. The final step in the initiation procedure is another rest period of 1 hour after which the HPPC pulse in Fig.\ref{fig:HPPC_Pulse_sim} is applied. After each HPPC pulse, the battery is discharged with a C/3 current to reduce the SOC by 10\%. The current profile for HPPC pulse (green) described in Section \ref{sec:hppcpulsedesc} and discharge to the next SOC is shown in Fig.\ref{fig:capdef}.

From Fig.\ref{fig:capdef}, the total coulombs to be removed in Ampere-seconds (\rm As) is written as
\begin{align}
360C  =  C_{\rm pulse} + C_{\rm rem}  = C(t_{d} - 0.75 t_{c}) + C_{\rm rem}  \label{Eqn:Cap_rem2} 
\end{align}
where $C_{\rm pulse}$ is the coulombs removed by the HPPC pulse and  $C_{\rm rem}$ denotes the coulombs to be removed by the discharge current of C/3 amperes in \rm As.
Here, the HPPC pulse current consists of the discharge pulse $C$ amperes for $t_{d}$ seconds and a charge pulse of current 0.75$C$ amperes for $t_{c}$ seconds. For the pulse in Fig.\ref{fig:HPPC_Pulse_sim}, the coulombs removed is written as
\begin{align}
	C_{\rm pulse}  &=  30C - 7.5 C = 22.5C  \quad {\rm As} \label{eq:Cpulse}
\end{align}
Substituting \eqref{eq:Cpulse} in \eqref{Eqn:Cap_rem2}, we get
\begin{align}
t_{\rm dis} = \frac{C_{\rm rem}}{C/3}	 =  \frac{337.5C}{C/3} = 1012.5{\rm s} = 16.875 \, {\rm minutes}
\end{align}
At the end of the test, the battery is fully discharged and the battery is rested. If $V_{\rm min}$ is reached during the HPPC pulse, the current is tapered to finish the pulse and if $V_{\rm min}$ is reached during C/3 discharge, the test is stopped.
The HPPC test procedure is summarized in Algorithm \ref{Alg:HPPC_method}.

\begin{figure}[h!]
	\begin{center}
		{\includegraphics[width= 1\columnwidth]{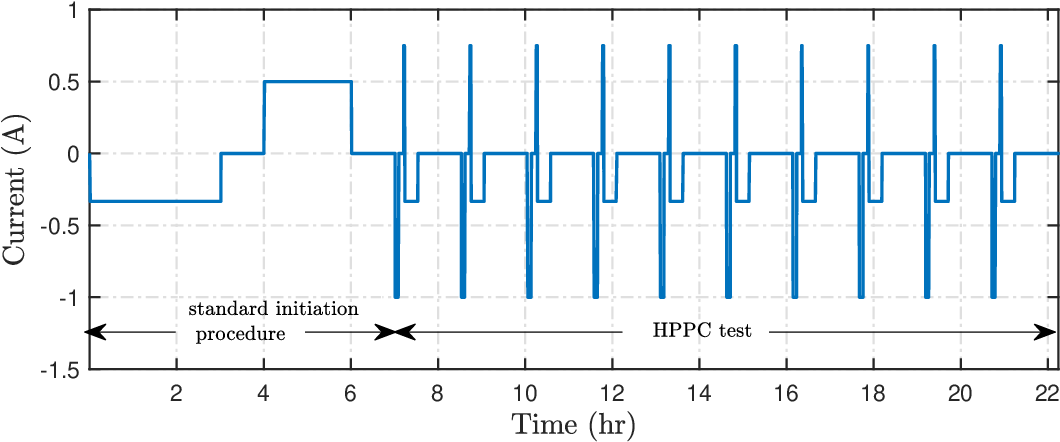}} \hspace{0.5cm}
		\caption{Full HPPC test current profile.}
		\label{fig:HPPC_Test}
	\end{center}
\end{figure}

\begin{algorithm}[h!]
	\caption{\bf{HPPC test procedure} ($C, I_{\rm taper}$ )\label{Alg:HPPC_method}}
	\begin{algorithmic}[1]
	
	\State{Discharge using  $I_{\rm dis} = C/3$ current until $v = V_{\rm min}$.}
	\State{Rest the battery for 1 hour.}
	\State{Charge the battery until $v = V_{\rm max}$} using standard charging process specified by the manufacturer.
	\State{Rest the battery for 1 hour.}	
	\State{Apply HPPC pulse profile described in Fig. \ref{fig:HPPC_Pulse_sim} of Section \ref{sec:hppcpulsedesc}.}
	\State{Discharge the battery by 10\% using $C/3 $ current.}
	\State{Rest the battery for 1 hour.}
	\State{Repeat steps 5-7 ten times until the battery is fully discharged.}
	
	\State{End test}
	\end{algorithmic}
	\end{algorithm}

In existing works \cite{christophersen2015battery,lee2015maximum}, the discharge and regenerative (charge) resistances,  $R_{\rm dis}$ and $R_{\rm regen}$ respectively, are calculated using the measured current and voltage values as
\begin{align}
	R_0 \triangleq R_{\rm dis}= \left|\dfrac{v(t_1)-v(t_0)} {i(t_1)-i(t_0)}\right|
	\label{DischargeResistance}
\end{align}   
and
\begin{align}
	R_{\rm regen} =\left|\dfrac{v (t_3)-v (t_2)}{i(t_3)-i(t_2)}\right|
	\label{ChargeResistance}
\end{align}

\section{Problem Definition}
\label{sec:problemDef}  

\begin{figure}[h!]
\begin{center}
{\begin{circuitikz}[american,scale=1.2, voltage dir = EF,line width=1pt]
\draw
(0,0) to [battery1, v= $\rE(s(k))$] (0,2)
(0,2) -- (0.5,2)
(0.5,2) -- (1,2)
(1,2) to[/tikz/circuitikz/bipoles/length=1cm,R=$ R_0$] (3,2) 
(4,2) to[short, i^=${i(k)}$, o-] (3,2)
(4,2) to[open, v=${v(k)}$] (4,0)
(4,0) to[short, o-] (0,0);
\end{circuitikz}}
\end{center}
\caption{R-int equivalent circuit model of a battery.}
\label{fig:RintECM}
\end{figure}
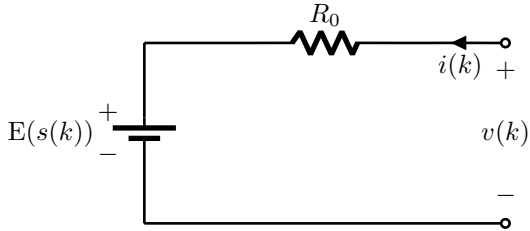

In this section, the limitation of the existing approach to resistance estimation is elaborated using the R-int equivalent circuit model of a battery shown in Fig.\ref{fig:RintECM}. 

The terminal voltage is given as
\begin{align}
v(k) = \rE(s(k)) + i(k) R_0
\end{align}
where 
$k$ denotes the time instance,
$\rE(s(k))$ is the open-circuit voltage at SOC $s(k)$, $i(k)$ is the current during discharging, and $R_0$ is the internal resistance of the battery.
Now, at time $t_0$, i.e., when the current is zero before the discharge pulse is applied, the terminal voltage is
\begin{align}
v(t_0) = \rE(s(t_0)) + i(t_0) R_0 = \rE(s(t_0)) 
\end{align}
where $t_0=0.$
At time $t_1$, i.e., at the end of the discharge pulse, the terminal voltage is
\begin{align}
v(t_1) = \rE(s(t_1)) + I_{\rm dis} R_0 \label{eq:voltdiff}
\end{align} 

The internal resistance of the battery can therefore be computed using
\begin{align}
v(t_1) - v(t_0) &=  \rE(s(t_1)) + I_{\rm dis} R_0 - \rE(s(t_0)) \\
v(t_1) - v(t_0) &=  \rE(s(t_1)) - \rE(s(t_0)) + I_{\rm dis} R_0 \\
\Delta v &=  \Delta \rE + I_{\rm dis} R_0 \label{eq:edefn} \\
R_0 &= \left|\frac{\Delta v -  \Delta \rE}{ I_{\rm dis}} \right|\label{eq:edefn2}
\end{align}
where $\Delta v = v(t_1) - v(t_0)$ and $\Delta \rE =  \rE(s(t_1)) - \rE(s(t_0))  $ denote the changes in the terminal voltage and OCV, respectively, between times $t_1$ and $t_0$.

Present resistance estimation approaches ignore $\Delta \rE$ and compute the internal resistance as 
\begin{align}
\hat R_0 = \left|\frac{\Delta v}{I_{\rm dis}} \right| \label{eq:hppctest}
\end{align}

\begin{remark}
In addition to the effect of EMF, the batteries are also subjected to the relaxation effect. 
Typically, RC circuits are employed to account for the relaxation effect. 
However, estimating the parameters of the RC circuits involves non-linear models and computationally intensive measurement and computational procedures. 
This is avoided in practical systems by standardizing the time difference between $t_0$ and $t_1$ \cite{barai2018study}.   
\end{remark}

The HPPC discharge pulse is 30 seconds long and the typical discharge current varies from 1C to 15C.
The high current magnitude and longer discharge duration will result in a significant OCV drop. 
Thus, the resistance computed using the HPPC test is overestimated by ($\Delta \rE/I_{\rm dis}$).
In the next section, a method to correct for the overestimation using an estimate of $\Delta \rE$ is proposed.

\section{Proposed Approach}
\label{sec:proposed}

In this section, an approach utilizing a novel observation model proposed in \cite{pillai2024open} is presented to improve the resistance estimates summarized in Section \ref{sec:problemDef}.
First, the slope of the OCV (or the EMF voltage) between times $t_0$ and $t_1$ is written as
\begin{align}
f'_{\rm em}\left\{s(t_1, t_0)\right\} =  \frac{\rE(s(t_1)) - \rE(s(t_0))}{ s(t_1) - s(t_0)} = \frac{\Delta \rE}{\Delta s} \label{eq:slopeOCV}
\end{align}
Here 
$s(t_0) $ is the SOC of the battery at time $t_0$,
$s(t_1) $ is the SOC of the battery at time $t_1$, and 
$\Delta s = s(t_1) - s(t_0)$ denotes the change in SOC. 
Similarly,
$\rE(s(t_0))$ is the OCV of the battery at time $t_0$, 
$\rE(s(t_1)) $ is the OCV of the battery at time $t_1$, and
$\Delta \rE = \rE(s(t_1)) - \rE(s(t_0))$ denotes the change in OCV. 

From \eqref{eq:slopeOCV}, the expression for the change in OCV between times $t_0$ and $t_1$ is written as
\begin{align}
\Delta \rE &=  f'_{\rm em}\left\{s(t_1, t_0)\right\}  \Delta s \label{eq:delRE}
\end{align}
Using rectangular approximation, the change in SOC is written as
\begin{align}
\Delta s  = \frac{T_s a(t_1, t_0)  I_{\rm dis}}{Q} = \frac{C\left\{a(t_1, t_0)\right\}}{Q}\label{eq:delS}
\end{align}
where $T_s$ is the sampling time, $a(t_1, t_0)$ is the number of windows given as
\begin{align}
a(t_1, t_0) = \frac{t_1-t_0}{T_s}
\end{align} 
and $C\left\{a(t_1, t_0)\right\} = T_s a(t_1, t_0)  I_{\rm dis}$ denotes Coulombs drawn between times $t_0$ and $t_1$ for a battery with capacity $Q$ at constant current $I_{\rm dis}$ and constant sampling time $T_s$.
The expression for the change in OCV in \eqref{eq:delRE} is rewritten as 
\begin{align}
\Delta \rE  &=  \frac{f'_{\rm em}\left\{s(t_1, t_0) \right\}}{Q}  C\left\{a(t_1, t_0)\right\} \\ &=  \kappa(t_1, t_0)  C\left\{a(t_1, t_0)\right\} \label{eq:kappaOCV}
\end{align}
where $ \kappa(t_1, t_0) =\frac{f'_{\rm em} \left\{s(t_1, t_0) \right\}}{Q}$.

The measured terminal voltage at time $t_0$, i.e., when the current is zero, is written as 
\begin{align}
v(t_0) = i(t_0) R_0 +  \rE(s(t_0)) =  \rE(s(t_0))  \label{eq:timet0}
\end{align}
At time $t_1$, the measured terminal voltage is
\begin{align}
v(t_1) =  i(t_1) R_0 +  \rE(s(t_1)) = I_{\rm dis} R_0 +  \rE(s(t_1)) 
\end{align}
Using \eqref{eq:slopeOCV} and \eqref{eq:kappaOCV}, the terminal voltage is rewritten as
\begin{align}
v(t_1) &=  I_{\rm dis} R_0 +  \rE(s(t_0)) +  \Delta \rE \\
&=  I_{\rm dis} R_0 +  \rE(s(t_0)) + \kappa(t_1, t_0)  C\left\{a(t_1, t_0)\right\} \label{eq:timet1}
\end{align}
Similarly, at any time $t_x \in [t_0, t_1]$, the terminal voltage is written as 
\begin{align}
v(t_x) &=  I_{\rm dis} R_0 +  \rE(s(t_x)) \nonumber \\
&=  I_{\rm dis} R_0 +  \rE(s(t_0)) +  \kappa(t_x, t_0)  C\left\{a(t_x, t_0)\right\}  \label{eq:timetx}
\end{align}
where it must be noted that the OCV at time $t_x$ is written as 
\begin{align}
\rE(s(t_x)) = \rE(s(t_0)) +  \kappa(t_x, t_0)  C\left\{a(t_x, t_0)\right\} 
\end{align}
It is assumed that the gradient (slope) of the OCV-SOC curve between times $t_0$ and $t_1$ is constant, i.e.,
\begin{align}
 f'_{\rm em}(s) &= f'_{\rm em}\left\{s(t_1, t_0)\right\}  = \dots = f'_{\rm em}\left\{s(t_x, t_0)\right\} = \dots  \nonumber \\ 
& \quad = \dots =  f'_{\rm em}\left\{s(t_0+T_s, t_0)\right\} 
\label{eq:assumption}
\end{align}
Thus, with this assumption
\begin{align}
\kappa = \kappa(t_1, t_0) = \dots = \kappa(t_x, t_0) = \dots = \kappa(t_0+T_s, t_0) 
\end{align}
Thus, \eqref{eq:timet1} and  \eqref{eq:timetx} are rewritten using the linearity assumption as
\begin{align}
v(t_1) &=  I_{\rm dis} R_0 +  \rE(s(t_0)) + \kappa C\left\{a(t_1, t_0)\right\} \label{eq:timet1new} \\
v(t_x) &=  I_{\rm dis} R_0 +  \rE(s(t_0)) +  \kappa C\left\{a(t_x, t_0)\right\}  \label{eq:timetxnew}
\end{align}
Based on \eqref{eq:timet0}, \eqref{eq:timet1new}, and  \eqref{eq:timetxnew}, the following vector observation model can be written
\begin{align}
	\begin{bmatrix}
		v(t_0)  \\
		v(t_0+T_s)  \\
		\vdots \\
		v(t_x) \\
		\vdots \\
		v(t_1) 
	\end{bmatrix} \,
	=
	\begin{bmatrix}
		0 & 1 & 0 \\
		I_{\rm dis}  & 1 &  C\left\{a(t_0+T_s, t_0)\right\}  \\
		\vdots & \vdots & \vdots \\ 
		I_{\rm dis}  & 1 &  C\left\{a(t_x, t_0)\right\}  \\
		\vdots & \vdots & \vdots \\ 
		I_{\rm dis}  & 1 &  C\left\{a(t_1, t_0)\right\} 
	\end{bmatrix}
	\begin{bmatrix}
		R_0\\
		\rE(s(t_0))\\
		\kappa
	\end{bmatrix} \label{eq:obs2var}
\end{align}
In vector format \eqref{eq:obs2var} is written as
\begin{align}
	\bz = \bH \bx + \bn \label{eq:vecobs}
\end{align}
The parameters $\bx$ are obtained through the following constrained optimization problem
\begin{align}
&\hat \bx = \arg\max_{\bx}  || \bz - \bH \bx ||_2 \\
&\text{subject to } \bx \geq  0  \label{eq:constraint}
\end{align}
where
\begin{align}
\hat \bx_{\rm LS} =
\left[\hat R_{0,{\rm LS}} \quad		\hat \rE_{\rm LS} (s(t_0))\quad		\hat \kappa_{\rm LS} \right]^T \label{eq:lsest}
\end{align}
In this paper, MATLAB's `lsqnonneg' function is utilized to solve the above optimization problem. 
The required resistance $R_0$ estimate is thus given in \eqref{eq:lsest}.
This can be verified by applying the estimated correction $\Delta \hat \rE$ as
\begin{align}
	\hat R_0  &=  \frac{\Delta v -   \Delta \hat \rE}{ I_{\rm dis}} \\
	&=  \frac{\Delta v -  \hat \kappa_{\rm LS}  C\left\{a(t_1, t_0)\right\}}{I_{\rm dis} } \label{eq:updatedhpccR} 
\end{align}
where $ \hat \kappa_{\rm LS} $ is estimated in \eqref{eq:lsest}.

\begin{remark}
It must be noted that in an ideal scenario, the parameters $R_0$, $E(s(k))$, and $\kappa$ are all positive, because, the resistance of an electric circuit, the voltage of a battery, and the gradient of the OCV-SOC curve, which is monotonically non-decreasing with SOC, are all positive. 
However, the gradient of the OCV-SOC curve changes with time and this work assumes it to be a constant in \eqref{eq:assumption}, albeit for a short period of time. 
As a result, the constraint \eqref{eq:constraint} is needed to make sure the estimates are meaningful. 
Further, the accuracy of the estimates will suffer proportionately to the degree to which the assumption \eqref{eq:constraint} is violated. 
Practical results on this observation can be found in Section \ref{sec:simresults}. 
\end{remark}
The resistance $\hat R_0$ calculated in \eqref{eq:updatedhpccR} using proposed approach takes OCV drop $\Delta \hat{\rm E}$ into consideration, which helps in a more accurate resistance estimation.

\section{Simulation Results}
\label{sec:simresults}

In this section, the use of the existing and proposed approaches for OCV estimation is presented using simulated battery data.
The simulation analysis provides an objective performance evaluation metric to evaluate the performance gain of the proposed approach compared to the existing one. 

\subsection{Existing Approach}
\begin{figure}[h!]
\begin{center}
\subfloat[][Discharge pulse in the HPPC test, $i(k) = 15C_{\rm batt} = 22.5$A.\label{fig:ECM2I}]
{\includegraphics[width= 1\columnwidth]{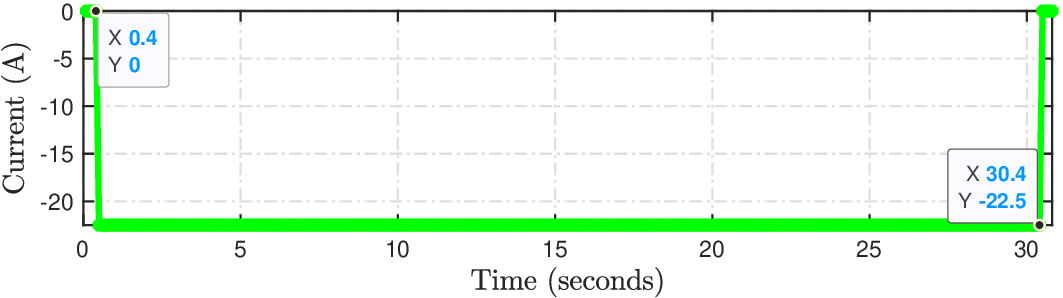}} \\
\subfloat[][Initial SOC, $s(t_0) = 1$.\label{fig:ECM2V}]
{\includegraphics[width= 1\columnwidth]{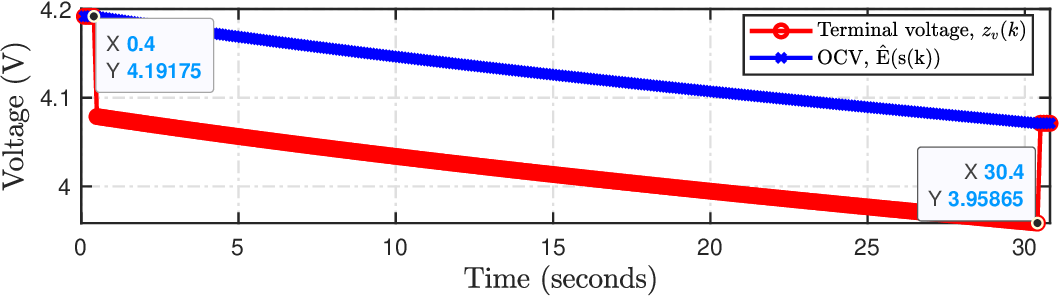}} \\
\subfloat[][Initial SOC, $s(t_0) = 0.5$.\label{fig:ECM2Vsoc0.5}]
{\includegraphics[width= 1\columnwidth]{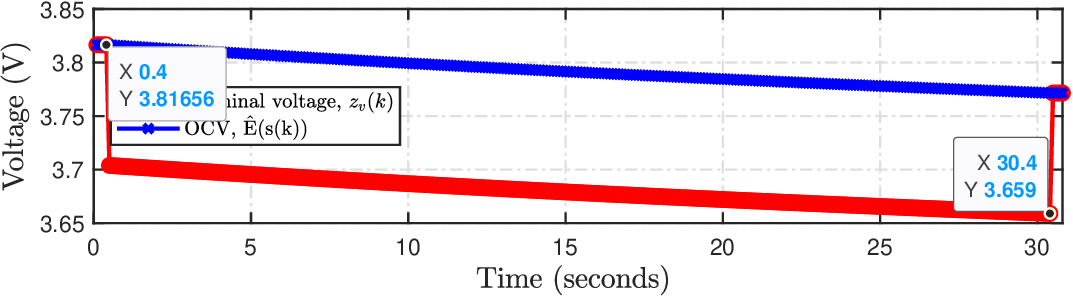}} \\
\subfloat[][Initial SOC, $s(t_0) = 0.15$.\label{fig:ECM2Vsoc0.15}]
{\includegraphics[width= 1\columnwidth]{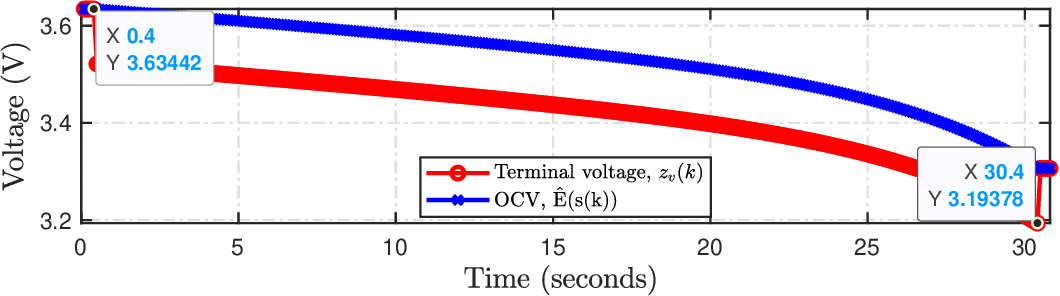}}
\caption{HPPC test current and voltage.\label{fig:ECM2VI}}
\end{center}
\end{figure}

For the simulation analysis, typical battery response is simulated using a battery simulator \cite{balasingam2023robust}. The options for the battery simulator are selected in such a way that it could be accurately represented using the ECM shown in Fig. \ref{fig:RintECM}.
 A typical discharge current pulse used in the HPPC test, the measured voltage $v(k)$ and the OCV $\hat \rE(s(k))$ were simulated as shown in Fig. \ref{fig:ECM2VI}. 

The true resistance used in the simulation was $R_0 = 5$m$\Omega$. The battery capacity is chosen to be $C_{\rm batt} = 1.5$Ah. It is assumed that the battery was full at the start, i.e., $s(t_0) = 1$. The OCV of the battery in Fig.\ref{fig:ECM2VI}\subref{fig:ECM2V}, in blue, was generated using the Combined+3 model \cite{pattipati2014open} that is given by
\begin{align}
{\rm E(k)} =& u_0 + \frac{u_1}{s(k)} + \frac{u_2}{s(k)^2} + \frac{u_3}{s(k)^3} + \frac{u_4}{s(k)^4} \nonumber \\
&+ {u_5}{s(k)} + u_6 \ln(s(k)) + u_7 \ln(1-s(k)) 
\label{eq:combined+3}
\end{align}
The generated OCV model had the following model parameters:
$u_0 = -9.082,$
$u_1 =  103.087,$
$u_2 =  -18.185,$
$u_3 =   2.062,$
$u_4 =   -0.102,$
$u_5 =  -76.604,$
$u_6 =  141.199,$ and
$u_7 =   -1.117$.

From Fig.\ref{fig:ECM2VI}\subref{fig:ECM2V}, the terminal voltage at $t_1 = 30.4$s is $v(t_1) = 3.9587$V and at $t_0 = 0.4$s, the terminal voltage is $v(t_0) = 4.1917$V.
Now, the resistance is computed using \eqref{eq:hppctest} as
\begin{align}
\hat R_0 &=\left| \frac{\Delta v}{I_{\rm dis}} \right| = \left| \frac{4.1917 - 3.9587}{-22.5} \right| \\
&= 10.3546 \, \text{m}\Omega 
\end{align}
The computed resistance based on existing aprocach is thus an over-estimated value compared with the true resistance due to changes in the OCV, i.e., $\hat R_0 >> R_0$.

Further, the variation in the initial SOC of the battery also affects the computed resistance. For example, the initial SOC values are chosen to be 0.5 and 0.15 in Figures \ref{fig:ECM2VI}\subref{fig:ECM2Vsoc0.5} and  \ref{fig:ECM2VI}\subref{fig:ECM2Vsoc0.15} respectively. Using \eqref{eq:hppctest}, the resistances computed are $7.0025$m$\Omega$ and $19.5838$m$\Omega$, respectively. 

 As discussed in Section \ref{sec:proposed}, the overestimation of resistance is due to OCV drop omission. For example, the value of $v(t_0)$ is 4.19175$\rm V$, when the current is 0A, i.e., at the start of the pulse in Fig.\ref{fig:ECM2VI}\subref{fig:ECM2V}. The presence of OCV prevents the terminal voltage from returning to its original value of 4.19175$\rm V$ at $v(t_0)$, i.e., when the current shifts to 0$\rm A$ again at the end of the pulse. The same pattern can be observed at other SOCs as shown in Figures \ref{fig:ECM2VI}\subref{fig:ECM2Vsoc0.5}, and \ref{fig:ECM2VI}\subref{fig:ECM2Vsoc0.15}. Since the OCV drop is neglected as shown in \eqref{eq:hppctest}, the computed resistance will be erroneous. Now, we define the error in the calculated resistance as 
\begin{align}
\epsilon_{R_0} &= \frac{| \hat R_0  - R_0 | }{R_0}  \times 100 \label{eq:errorR}
\end{align}
Table \ref{tabl:Rsoc} summarizes the terminal voltages and the corresponding resistance computed for three different initial SOC values. In Table \ref{tabl:Rsoc}, the error in resistance estimate is shown in the last row as a percentage. It is further confirmed from the percentage values that the resistance is over-estimated by a minimum of 40\% when SOC is 50\%. A maximum of 291\% error in the resistance is observed in low-SOC regions.  

\begin{table}[h!]
\begin{center}
\caption{Computed resistances using the existing approach at three different initial SOC.\label{tabl:Rsoc}}
\begin{tabular}{|c|c|c|c|}
\hline
$s(t_0)$         & 1  & 0.5 & 0.15    \\ \hline
$v(t_0)$ (V)  & 4.1917 &  3.8166 & 3.6344   \\ \hline
$v(t_1)$ (V)  & 3.9586  & 3.6590 & 3.1938   \\ \hline \hline
$I_{\rm dis}$ (A)  & -22.5 & -22.5  &  -22.5  \\ \hline \hline
%
%
$R_0({\rm HPPC})$ (m$\Omega$) & 10.36 &  7.0025 & 19.5838  \\ \hline
$\epsilon_{R_0}$ ($\%$) & 107.2003 &  40.0501 & 291.6755   \\ \hline
\end{tabular}
\end{center}
\end{table}

\begin{figure}[h!]
\begin{center}
{\includegraphics[width= 1\columnwidth]{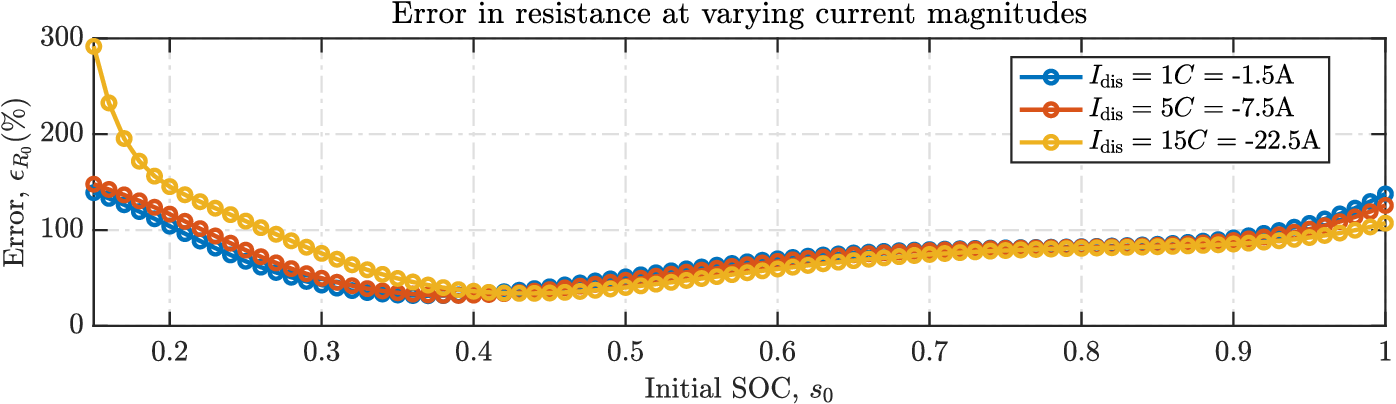}}
\caption{Error in resistance computed using the existing approach over various initial SOC values, $s(t_0) = 0.15$ to $1$ for three current discharge settings.}
\label{fig:ECM2R}
\end{center}
\end{figure}

Fig.\ref{fig:ECM2R} shows the error in resistances (in yellow) over the SOC regions ranging from $s(t_0) = 0.15$ to $1$ for a current magnitude of $22.5$A. It can be observed that the resistance is severely overestimated in the lower-SOC regions. In other regions, the resistance is over-estimated by at least 30$\%$. 
Further, at lower current magnitudes of $1.5$A and $7.5$A in Fig.\ref{fig:ECM2R}, the errors are lower but substantial and are in the range (0\%,140\%) of the assumed true resistance value of $5$m$\Omega$. In Fig.\ref{fig:ECM2R}, the highest error is in the range of $0$ to $300$\% and is observed at a higher current of $22.5$A.

\begin{remark}
It must be noted that in an actual battery, the polarization effect may further deteriorate the estimation performance.
In the simulation studies, only the Ohmic resistance was considered to objectively demonstrate the performance gain by the proposed approach through a controlled experiment. 
\end{remark}

\subsection{Analysis of the Proposed Approach}

Table \ref{tabl:Rsocpa} presents the resistance values computed using the corrected terminal voltage in \eqref{eq:updatedhpccR} at three different initial SOC levels. The table also shows the resistance error as a percentage computed using \eqref{eq:errorR}. From the table, it can be observed that a maximum of 49.3118$\%$ error in the resistance is observed at low-SOC regions. 


\begin{table}[h!]
\begin{center}
\caption{Computed resistances using corrected terminal voltage and current at times $t_0$ and $t_1$ for different initial SOC.\label{tabl:Rsocpa}}
\begin{tabular}{|c|c|c|c|}
\hline
$s(t_0)$         & 1  & 0.5 & 0.15    \\ \hline
$\hat v(t_0)$ (V)  & 4.1917 &  3.8166 & 3.6344   \\ \hline
$\hat v(t_1)$ (V)  & 4.0782  & 3.7037 & 3.5209 \\ \hline \hline
$I_{\rm dis}$ (A)  & -22.5 & -22.5  &  -22.5  \\ \hline \hline
$\hat \rE(t_0)$ (V)  & 4.1917 &  3.8166 & 3.6344 \\ \hline
$\hat \rE(t_1)$ (V)  & 4.0729 & 3.7719  & 3.3618   \\ \hline 
$\rm \Delta \hat E$ (V)  &0.1188   & 0.0446  & 0.2727   \\ \hline \hline
$\hat R_0$ (m$\Omega$) & 5.0784  & 5.0193 & 7.4656  \\ \hline
$\epsilon_{R_0}$ ($\%$) &  1.5678   &  0.3867 & 49.3118  \\ \hline
\end{tabular}
\end{center}
\end{table}

\begin{figure}[h!]
\begin{center}
{\includegraphics[width= 1\columnwidth]{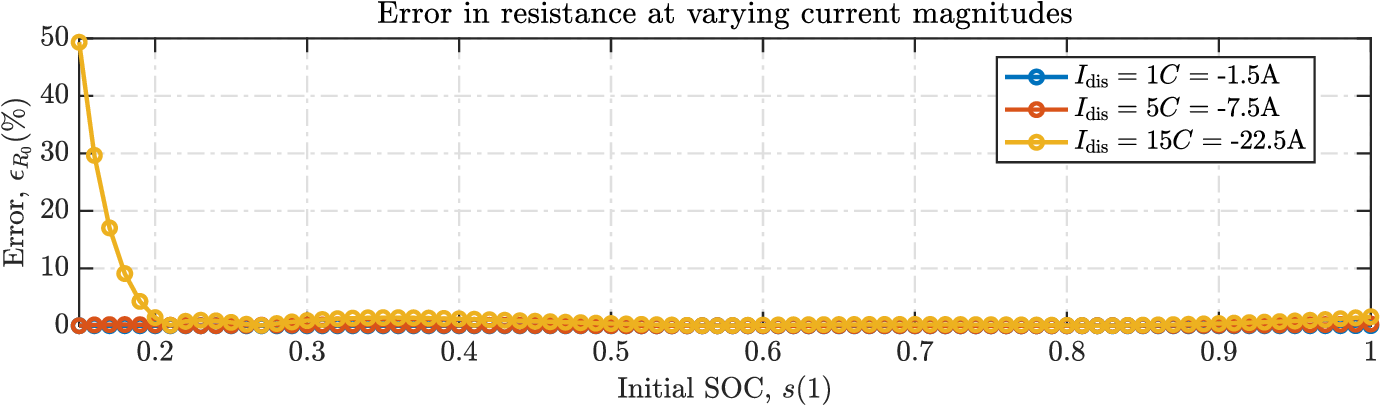}}  
\caption{Error in resistance computed using the proposed approach over various initial SOC values, $s(t_0) = 0.15$ to $1$ for three current variations.}
\label{fig:ECM2Rpa}
\end{center}
\end{figure}

In Fig.\ref{fig:ECM2Rpa}, the error in resistance estimates (in red) for SOC ranges from $s(t_0) = 0.15$ to $1$ is shown. It can be observed that lower errors are observed especially when the current is lower. At a higher current of $22.5$A, higher errors in the resistance estimates are observed.

\subsection{Comparison of the proposed and existing approaches}

\begin{figure}[h!]
\begin{center}
\subfloat[][\label{fig:ECM2Vsoc0.5Comp}]
{\includegraphics[width= 1\columnwidth]{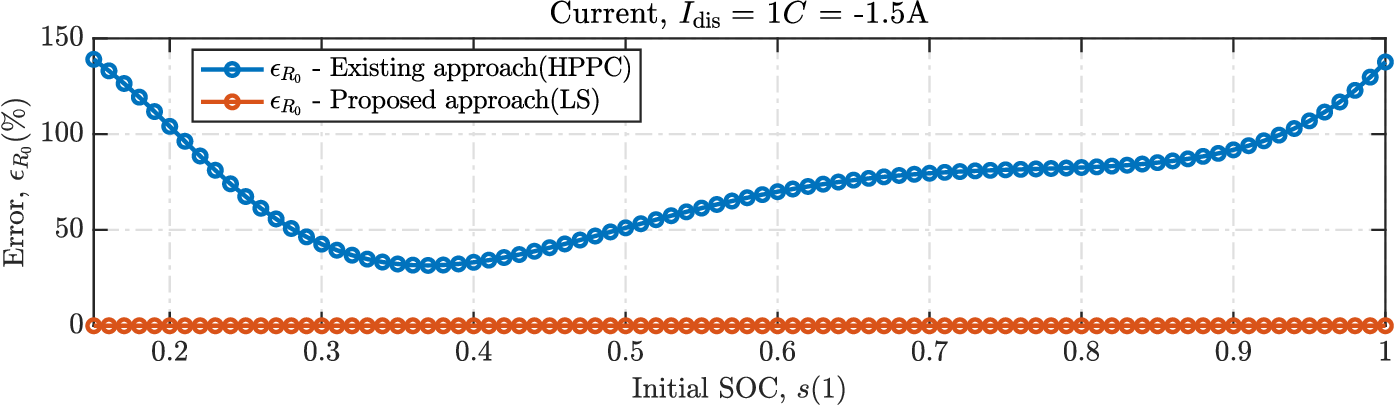}}  \\
\subfloat[][\label{fig:ECM2Vsoc0.1Comp}]
{\includegraphics[width= 1\columnwidth]{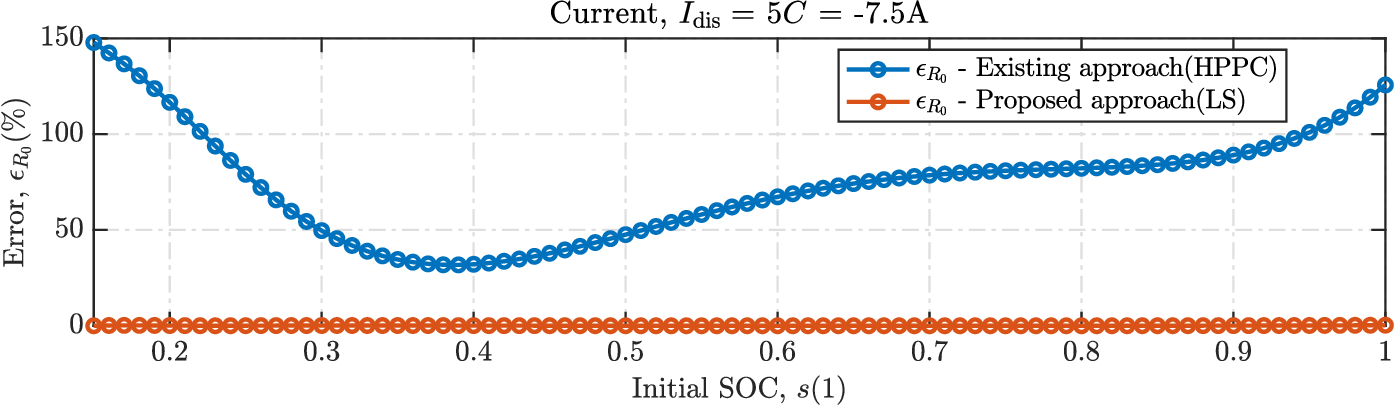}}  \\
\subfloat[][\label{fig:ECM2Vsoc1Comp}]
{\includegraphics[width= 1\columnwidth]{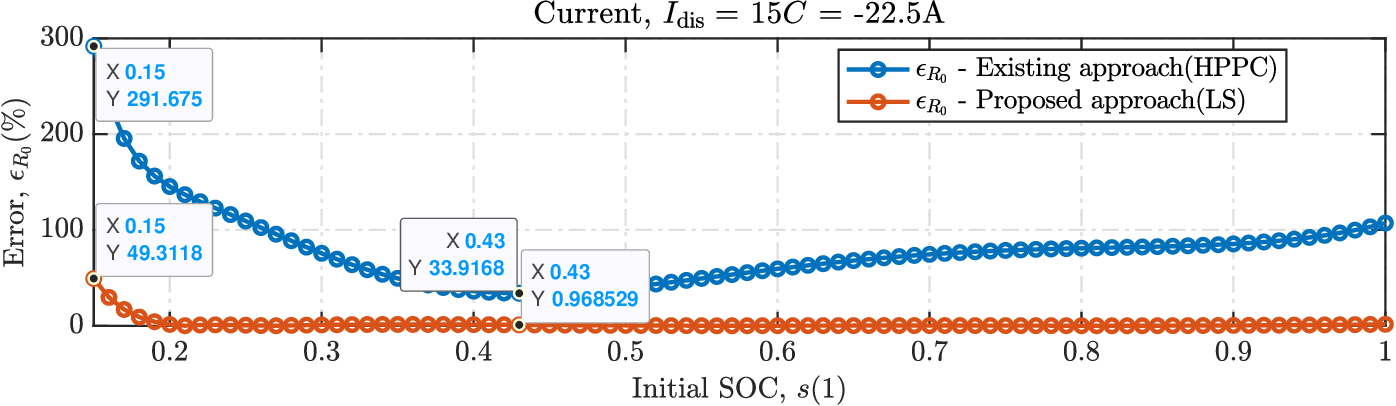}} 
\caption{Existing (HPPC) and proposed approaches' resistance estimation over various initial SOC values, $s(t_0) = 0.15$ to $1$ for three current variations.}
\label{fig:ECM2RComp}
\end{center}
\end{figure}

In this subsection, the performance of the proposed and existing (HPPC) approaches are compared using the error values computed using \eqref{eq:errorR}. 
Fig.\ref{fig:ECM2RComp} shows the errors in resistance estimates of the existing (in blue) and proposed (in red) for three current magnitudes of $1C$, $5C$, and $15C$.
It can be noticed that the error is reduced when the proposed approach to OCV estimation is used as part of resistance estimation at all current values. As expected, lower errors are observed at lower currents in both approaches as seen in Figures \ref{fig:ECM2RComp}\subref{fig:ECM2Vsoc0.5Comp} and \ref{fig:ECM2RComp}\subref{fig:ECM2Vsoc0.1Comp}. The highest error is observed at the maximum current magnitude of $22.5$A in Fig.\ref{fig:ECM2RComp}\subref{fig:ECM2Vsoc1Comp}. Even at higher current magnitudes, a lower error is observed for the proposed approach when compared to the existing approach to resistance estimation from the HPPC tests.

Now, a measure of performance gain as the maximum and minimum difference between the previous and proposed approach is defined. From Figure \ref{fig:ECM2RComp}, the maximum difference, termed gain of the proposed approach, is at the higher current magnitude of $22.5$A at $291.67-49.31 = 242.36\%$.
The minimum gain was observed to be similar across all current magnitudes at $33-0.96 = 32.94\%$.

\section{Experimental Results}
\label{sec:data-collection}

In this section, the proposed approach is demonstrated using data collected from cylindrical Li-ion batteries. 

\subsection{Data collection setup}
\label{sec:arbincycler}

\begin{figure}[h!]
\begin{center}
{\includegraphics[width= .4\columnwidth]{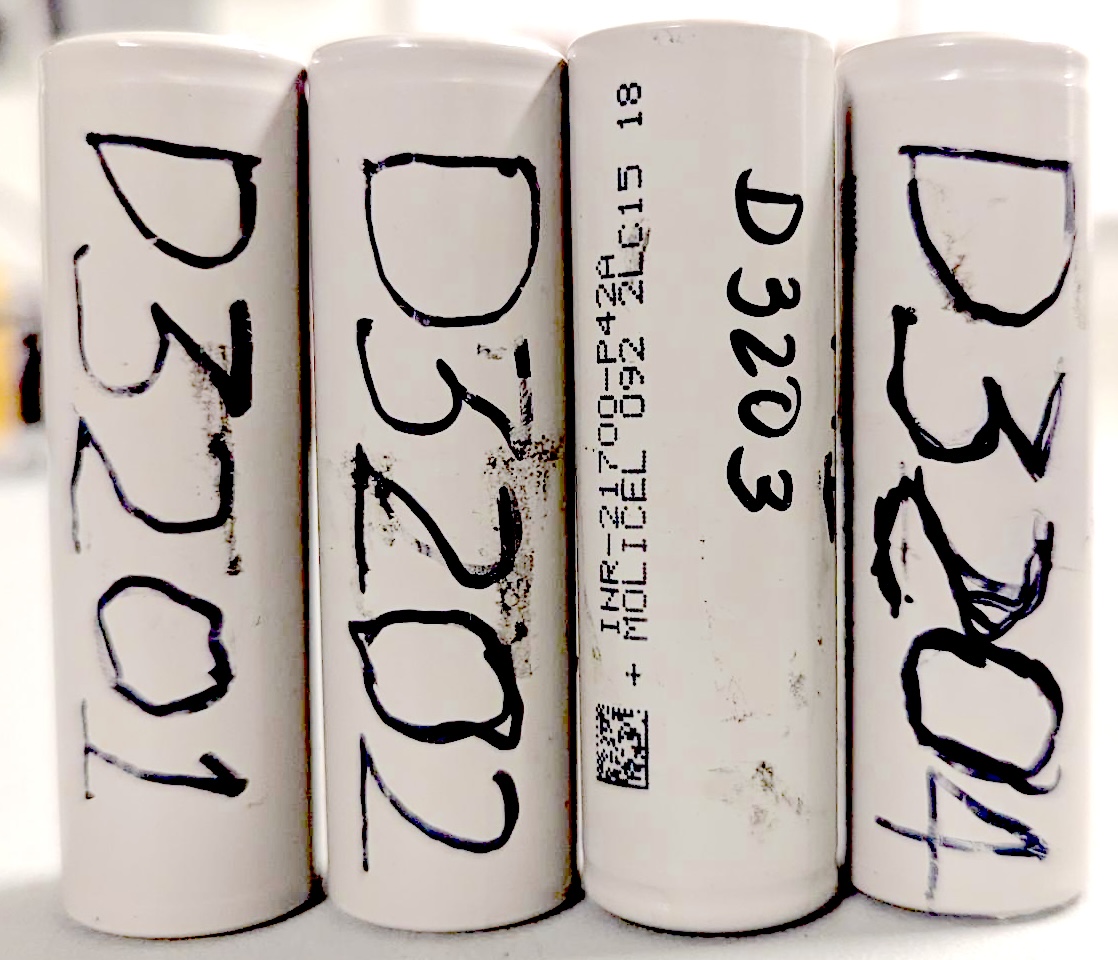}} \hspace{0.2cm}
\caption{Molicel INR-21700-P42A batteries.}
\label{fig:molicel}
\end{center}
\end{figure}

Voltage, current and time data for this study were collected using the Laboratory Battery Test (LBT) Series LBT-21084 model at the Battery Management Systems Lab (BMSLab) of the University of Windsor \cite{arbin_battery_test_equipment_2024}. 
Four Molicel INR-21700-P42A cells \cite{molicel_inr21700p42a_2024} were used in this study.
A photograph of the four cells labeled as D3201, D3202, D3203, and D3204, used in the experiment is shown in Fig.\ref{fig:molicel}. Table \ref{tabl:molicel} provides a summary of features of the battery model from its datasheet. 

\begin{table}[h!]
\caption{{Molicel INR-21700-P42A battery specifications.}\label{table:data-collection-plan}}
\label{tabl:molicel}
\begin{center}
\begin{tabular}{|c|c|} \hline
\textbf{Specification}               & \textbf{Battery} \\ \hline
Nominal voltage                      & $3.6$V           \\ \hline
Minimum capacity, $C$                & $4.2$ Ah       \\ \hline
Discharge current                    & $45$A            \\ \hline
Height                               & $70.2$ mm        \\ \hline
Diameter                             & $21.7$ mm        \\ \hline
Weight                               & $70$g            \\ \hline
Internal resistance                  & $16$ m$\Omega$   \\ \hline
Maximum voltage, $ V_{\rm max}$ & 4.2 V            \\ \hline
Minimum voltage, $ V_{\rm min} $ & 2.5 V            \\ \hline
\end{tabular}
\end{center}
\end{table}

In this data collection, the protocol in Algorithm \ref{Alg:HPPC_method} was implemented using the Arbin cycler \cite{arbin_battery_test_equipment_2024}. 
The two inputs for the algorithm were chosen as $C=4.2$Ah, and the taper current for the CC-CV charging protocol was chosen as $I_{\rm taper}=20$mA. 
The discharge current used for the HPPC pulse was $1C = 4.2 \rm A$ and sampled at a frequency of $1$ Hz. The procedure is repeated for all four batteries and the current, voltage and time data were recorded for further analysis. The next section describes the recorded data and the results of resistance estimation on this data. 

\subsection{Data Analysis}

\begin{figure}[h!]
\begin{center}
\subfloat[][HPPC test current.\label{subfig:CurrentHPPCpulse}]
{\includegraphics[width=.99\columnwidth]{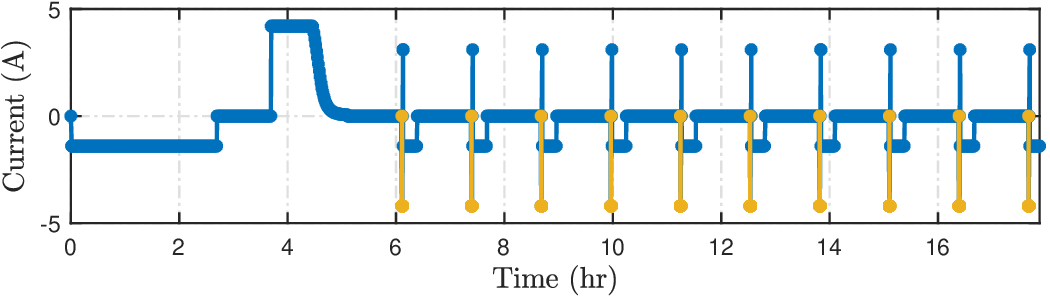}} \\
\subfloat[][HPPC test voltage.\label{subfig:VoltageHPPCpulse}]
{\includegraphics[width=.99\columnwidth]{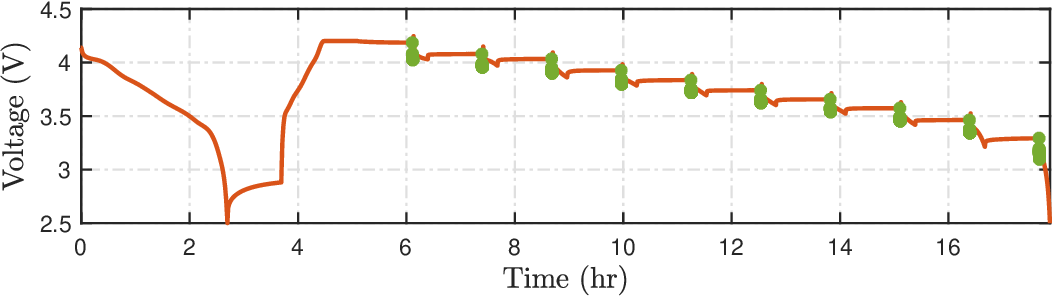}} \\
\subfloat[][HPPC pulse current.\label{subfig:CurrentHPPCpulseSS}]
{\includegraphics[width=.5\columnwidth]{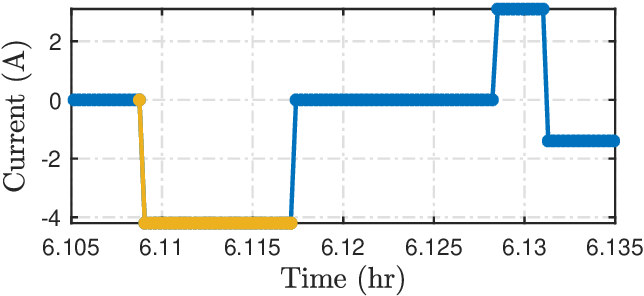}} 
\subfloat[][HPPC pulse voltage.\label{subfig:VoltageHPPCpulseSS}]
{\includegraphics[width=.5\columnwidth]{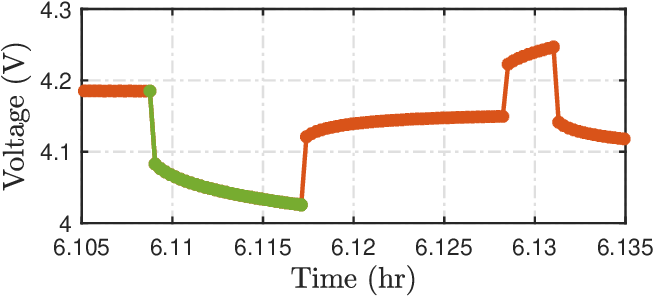}} 
\caption{HPPC test current and voltage data from Molicel INR-21700-P42A battery.\label{fig:VIhppc}}
\end{center}
\end{figure}

The data of interest in this study were the voltage, current and time recorded during the HPPC test in Fig.\ref{fig:VIhppc}. The current recorded during the HPPC test is shown in Fig.\ref{fig:VIhppc}\subref{subfig:CurrentHPPCpulse} and the voltage measurements in Fig.\ref{fig:VIhppc}\subref{subfig:VoltageHPPCpulse}. The discharge pulses were extracted from these measurements at different SOC levels to demonstrate the proposed approach to resistance estimation from Section \ref{sec:proposed}. The extracted discharge pulses are highlighted in yellow and green in Fig.\ref{fig:VIhppc}. An enlarged version of one discharge pulse and its extracted current and voltage measurements are shown in Figures \ref{fig:VIhppc}\subref{subfig:CurrentHPPCpulseSS} and \ref{fig:VIhppc}\subref{subfig:VoltageHPPCpulseSS}.

\begin{figure}[h!]
\begin{center}
{\includegraphics[width=.99\columnwidth]{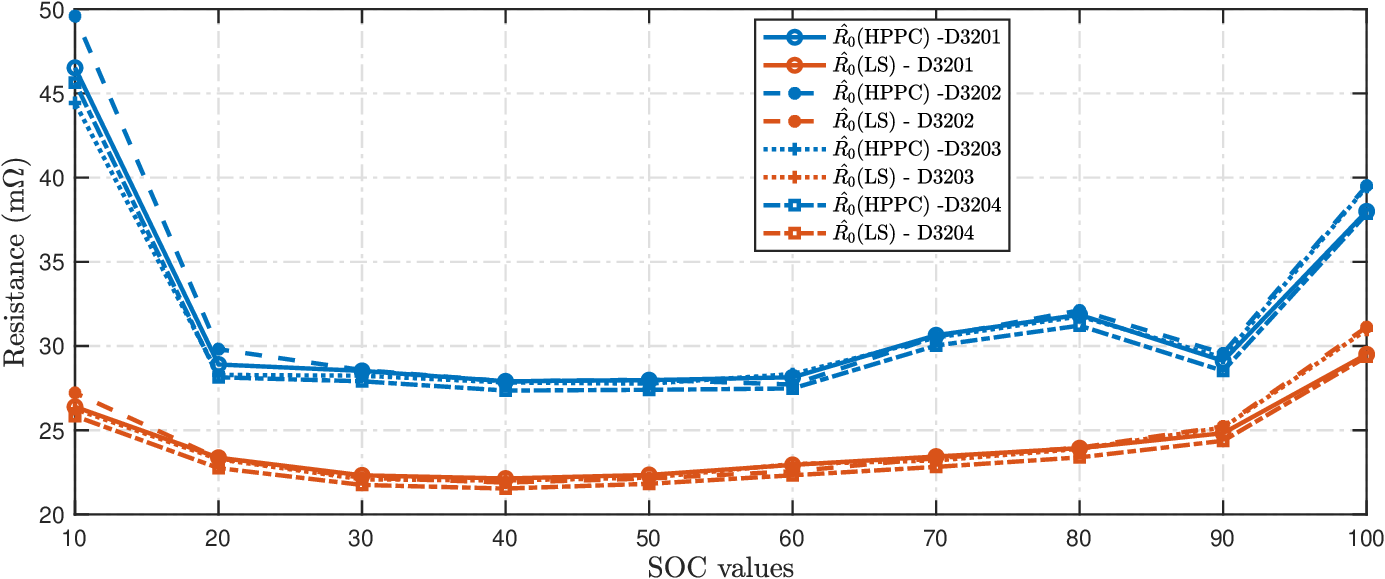}}
\caption{Resistance values computed for the previous(HPPC) and proposed approaches using real battery data from the Molicel INR-21700-P42A batteries.\label{fig:RMSEerrors}}
\end{center}
\end{figure}

The two methods of resistance estimation compared in this paper are presented in Fig.\ref{fig:RMSEerrors} from the recorded current and voltage measurements. The estimates computed using the HPPC standard are denoted as $\hat R_0 ({\rm HPPC})$ and the estimates from the proposed approach in Section \ref{sec:proposed} are denoted as $\hat R_0 $. The computation is repeated for data from four batteries as shown in the legends in Fig.\ref{fig:RMSEerrors}.

A comparison of resistance estimates using the two methods shows that the existing approach overestimated the resistance as predicted for all SOC ranges. 
The overestimation of resistance is shown to be corrected using the proposed technique to account for the change in OCV when the discharge pulse is applied. 

%
%
%

\section{Conclusion}
\label{sec:conclusion}

Internal resistance estimation is crucial in many electronic systems, including the battery management system and the battery charger. 
Most accurate battery equivalent models include resistor-capacitor (RC) components to account for the relaxation effect of the battery.  
However, estimating the RC components will require precise measurement and extensive processing systems. 
In many practical systems, the battery internal resistance is estimated using a current excitation pulse of standardized duration. 
This paper presents an improved approach for such internal resistance estimation in batteries, particularly, in practical battery monitoring hardware where the internal resistance is instantly estimated as the ratio of the voltage drop and the applied current.

It is shown in this paper that existing approaches overestimate the internal resistance due to variations in the battery open circuit voltage (OCV) due to the applied excitation signal. 
It was further demonstrated that the amount of overestimation depends on the magnitude of the applied current excitation pulse and that the amount of such overestimation varies with the state of charge (SOC) of the battery. 
Further, an approach is presented in this paper to correct such overestimation.
The proposed correction approach utilizes a novel observation model to estimate the OCV drop without requiring any additional information about the battery, such as the OCV parameters, battery capacity, or the battery chemistry.
The proposed approach is universally applicable to batteries of any size, chemistry, and age. 

Realistic simulation analysis showed that the proposed approach can result in a performance gain from 30\% to 250\% in estimation error in standardized hybrid pulse power characterization (HPPC) tests. 
Further, it is reported through experimental validations that the proposed approach reduced the overestimate of the internal resistance of the batteries from 5 milli Ohms to 20 milli Ohms. The performance gains varied, as expected, based on the applied HPPC pulse current and the battery SOC.

A limitation of the proposed approach is that it assumes that the gradient of the OCV-SOC curve remains constant during the HPPC pulse. In reality, however, the gradient of the OCV-SOC curve changes with SOC. Particularly, the change of gradient is significant between about 10\% and 30\%  of the battery SOC. The accuracy of the proposed approach is shown to suffer in this range. To improve performance in this range, future works should consider higher-order models to model and estimate the OCV drop in this range.

\balance

\bibliographystyle{ieeetr}
\bibliography{LiteratureCollection}

\end{document}